\documentclass[preprint]{aastex}

\shortauthors{Wiebe \& Watson}
\shorttitle{Irregular  Magnetic  Fields  and the Linear Polarization of 
Starlight}

\begin{document}

\title{Irregular  Magnetic  Fields  in  Interstellar  Clouds  and
the Linear  Polarization  of  Starlight}

\author{Dmitri S. Wiebe\altaffilmark{1,2} and William D. Watson\altaffilmark{1}}
\affil{Department of Physics, University of Illinois, 1110 W. Green Street,
Urbana, IL 61801; dwiebe@pingora.physics.uiuc.edu; w-watson@uiuc.edu}
\altaffiltext{1}{also Department of Astronomy}
\altaffiltext{2}{Permanent address: Institute of Astronomy of the RAS,
48, Pyatnitskaya str., 109017 Moscow, Russia}

\begin{abstract}
Calculations are performed for the linear polarization of starlight due to 
extinction by
aligned dust grains when the starlight traverses a medium with irregular 
magnetic fields.
This medium is intended to represent the optically thick components of 
interstellar clouds
which are observed to make little, if any, contribution to the polarization of 
starlight. In
agreement with the observations, we find that the polarization properties of the
starlight---the average fractional polarization and the dispersion in position 
angles---can
be essentially unchanged. For this, the rms of the irregular component must be 
greater
than the average magnetic field, which in turn tends to imply that the turbulent 
velocities
in these interstellar clouds are super-Alfv\'enic.
\end{abstract}


\section{Introduction}

The fractional linear polarization of starlight tends to increase with 
extinction in the
dilute, interstellar medium (ISM) as a result of the selective extinction by 
non-spherical
dust grains which are aligned relative to the galactic magnetic field (e.g.,
Hiltner 1956). However, when the extinction occurs in the more dense medium of 
interstellar
clouds, the increase in polarization with increasing extinction is noticeably 
reduced
(Gerakines, Whittet, \& Lazarian 1995; Goodman et al. 1995; but cf. Hough et al. 
1988). When the
starlight traverses clouds with visual extinctions (magnitudes) $A_V\approx10$, 
the grains can
have little if any effect on the polarization characteristics of the starlight. 
The reduced
polarizing efficiency of grains has been observed to begin at an extinction
$A_V=1$ to 2 (Arce et al. 1998; Harjunp\"a\"a et al. 1999). This behavior has 
been
interpreted as most likely due to the failure of the alignment mechanisms for 
the grains
at the higher gas densities and optical depths. However, the widespread 
observation of
linear polarization in the {\em emission} at far IR and submillimeter 
wavelengths by the grains in even more
thick clouds (Hildebrand 1996) indicates that alignment mechanisms can be 
effective
at the higher gas densities, at least under certain conditions.

An alternative interpretation is that the grains are aligned with the magnetic 
field, but
that their contributions to the linear polarization nearly cancel because of 
rapidly changing
directions of the grain alignment along the line of sight due to irregularities 
in the
magnetic fields (e.g., Jones 1996).  This interpretation has seemed to be 
incompatible with the requirement
from the observations that both the fractional polarization and the dispersion 
in the
position angles of the polarization can be essentially unchanged when the 
radiation
traverses the cloud. We re-examine this conclusion by performing detailed 
calculations
for the linear polarization of radiation that passes through an idealized medium 
which
consists of dust grains that are aligned with irregular magnetic fields. 
Supersonic
turbulent velocities are common in interstellar clouds (e.g., Crutcher 1999) and 
tend to
indicate that turbulent magnetic fields must be present as well. Following Arons 
\& Max (1975), irregular magnetic fields
also have been widely discussed as playing a role in various aspects of the 
structure of
interstellar clouds and in star formation.

\section{Basic Methods}

Representative, irregular magnetic fields are created by statistical sampling of 
the
Fourier components of a power spectrum with the Kolmogorov (power law) form and 
with
Gaussian distributions for the amplitudes. These methods are standard (e.g.
Dubinski, Narayan, \& Phillips 1995) and have been
described in detail with our applications of them elsewhere (Watson, Wiebe, \& 
Crutcher
2001; also, Wallin, Watson, \& Wyld 1998). Based on the correlation function 
that we
compute from available MHD simulations (see below), we utilize a  power spectrum 
that
is somewhat steeper than Kolmogorov (wavenumber $k^{-7/3}$ instead of 
$k^{-5/3}$). Two
quantities enter to specify the magnetic fields created in this way---the rms 
value of
a spatial component of the irregular magnetic field $B_{\rm rms}$ and the 
correlation
length of these fields. We specify the latter in terms of the number of 
correlation
lengths $N_{\rm corr}$ along an edge of the cubic volume of computational grid 
points
in which the fields are created. A correlation length is specified here as the
separation at which the structure function of the magnetic fields
reaches essentially its asymptotic value (see Watson et al. 2001). Since the 
medium is
assumed to be isotropic, $B_{\rm rms}$ will be the same along any one of the 
three orthogonal
coordinate axes. The irregular magnetic field at each location is added to a 
constant,
average magnetic field $\mbox{\bfseries\itshape B}_{\rm avg}$ to create the 
total magnetic field.
We restrict our attention to directions for $\mbox{\bfseries\itshape B}_{\rm 
avg}$ that are
perpendicular to the line of sight; this component is of most importance for 
linear
polarization due to grains.

We also perform computations with magnetic fields that are the result of
time-independent, numerical simulations by others for compressible MHD 
turbulence. We
refer the reader to papers by these authors for a detailed description of the 
calculational
methods (see Stone, Ostriker, \& Gammie 1998). Certain basic aspects of the 
simulations
are summarized in our earlier paper (Watson et al. 2001). Magnetic fields from 
two
simulations will be utilized. In these simulations, the ratio $B_{\rm 
rms}/B_{\rm avg}$
is approximately 0.6 (medium) and 1.5 (weak), where the labels medium and weak 
refer to
the relative strength of the average magnetic field. Both simulations have 
approximately
the same supersonic turbulent velocities. The Mach number is about five. The 
ratio of the
turbulent velocity to the Alfv\'en velocity (the Alfv\'enic Mach number) is 1.6 
and 5,
respectively, in the medium and weak field cases.  For both simulations, we 
determine
$N_{\rm corr}$ to be about twelve. Comparisons of the polarization computed with 
these
MHD fields and with the
statistically created fields provide an indication of the uncertainty in the 
calculations of
the linear polarization to our approximate description of the irregular magnetic 
fields.
The MHD simulations also contain information on the variation of the matter 
density and
on its correlation with the variations in the magnetic field which are absent in 
our
statistically created medium. Creating the magnetic fields by statistical 
sampling does,
however, allow us to explore a wider range of values for $N_{\rm corr}$ and
$B_{\rm rms}/B_{\rm avg}$. The cubic volumes used in the computations have 256 
grid points
on a side.

Stokes $Q$ and $U$ intensities describe the linear polarization of the 
radiation.
When the fractional linear polarization $p\ll1$ as is the case here,
$Q$ and $U$ depend upon the integrals $q$ and $u$ where
\begin{equation}
q=\int\rho(s)\cos^2\gamma\cos2\phi\,{\rm d}s
\end{equation}
and the integral $u$ is obtained by replacing $\cos2\phi$ by $\sin2\phi$ in 
equation (1)
[e.g., Lee \& Draine 1985]. Here, the integration is along the distance $s$ of a
straight-line path of a ray of starlight, $\rho(s)$ is the matter density, and 
$\gamma$
is the angle between the magnetic field and the
plane of the sky. Except when we use the variations in density from the MHD 
simulations,
$\rho(s)$ is taken as a constant. The angle $\phi$ is the angle between the 
projection of the
magnetic field onto the plane of the sky and a reference
axis in this plane. Then
\begin{equation}
p=(p/A_V)_{\rm max}\,g\sqrt{q^2+u^2},
\end{equation}
where $g$ is the factor (approximated here as a constant in the ISM) that 
relates
the column density of matter to the optical extinction and $(p/A_V)_{\rm max}$ 
is the
maximum value of the ratio of the fractional linear polarization to the optical
extinction.  This maximum value is assumed to occur
toward sources where the magnetic field, and hence the direction of the 
alignment of
the dust grains, does not change along the line of sight. The position angle 
$\alpha$
of the linear polarization is ${\rm tan}^{-1}(u/q)/2$.

\section{Application to Interstellar Clouds}

The key observational information is that the characteristics of the linear 
polarization of
the starlight that passes only through the intercloud medium and the periphery 
of an ISM
cloud are observed to be essentially the same as the starlight that also passes 
through
the  more dense part of the cloud for clouds with $A_V$ of ten or so magnitudes. 
The
calculations here will
thus examine whether rays of starlight with a distribution of fractional 
polarizations and
position angles similar to what is observed for the starlight at the periphery 
of a cloud can
traverse a medium that is representative of the thick part of the cloud without 
these
polarization characteristics being altered significantly. We examine this issue 
by utilizing
the statistical sampling to create magnetic fields at the $(256)^3$ grid points 
in a cubic
volume to represent the dense component of the cloud (the ``dense cube''), and 
then
performing the integrals to obtain $Q$ and $U$ for the $(256)^2$ rays that 
propagate along the
grid lines of the cube. To create the $(256)^2$ rays with an initial  
distribution of
fractional
polarizations and position angles similar to that of the starlight which is 
observed at the
periphery of the cloud, we first allow the rays to propagate through another 
cubic volume
of magnetic fields (the ``diffuse cube''). The properties of this diffuse cube
($N_{\rm corr}$, $B_{\rm rms}/B_{\rm avg}$ and the extinction $A_V$ through the 
cube)
are chosen so that the rays
emerge from it with the desired polarization characteristics. The parameters 
that describe
the diffuse cube should not be viewed as additional ``free parameters'' since 
this cube is
only a device for creating a distribution of rays with polarization 
characteristics similar to
what is observed at the periphery of the cloud. It is noteworthy that choices 
for the
diffuse cube which lead to rays with the desired
average fractional polarization and dispersion in position angles are
$B_{\rm rms}/B_{\rm avg}=0.6$ as inferred for the general ISM and the reasonable 
value
$N_{\rm corr}=4.5$ (e.g., Jones, Klebe, \& Dickey 1992).

The behavior of the average fractional polarization $\bar{p}_{\rm K}$ and the 
rms
dispersion angle $\sigma_\alpha$ as the rays traverse the dense cube is shown in 
Figure 1.
Since the observations are at K band, $(p_{\rm K}/A_V)_{\rm max}=0.64$~\% 
mag$^{-1}$
is used in equation (2). This value is computed from Serkowski (1973) law
and multiplied by 1.4 from Mathis (1986).
In the observations with which we are
comparing, $\bar{p}_{\rm K}$ typically is about one percent and $\sigma_\alpha$
is 10--20 degrees for the starlight at the periphery of the cloud (Goodman et 
al. 1995).
When the light that enters the cloud already is polarized,
it is clear from Figure 1 that the change in $\bar{p}_{\rm K}$ and 
$\sigma_\alpha$ can be
quite small if there are a number ($\ga12$) of correlation lengths across the 
dense cube and
if $B_{\rm rms}$ is somewhat greater than $B_{\rm avg}$. The observational data
reproduced in Figure 1 exhibit considerable scatter away from any single curve
associated with a specific $N_{\rm corr}$ and $B_{\rm rms}/B_{\rm avg}$. This is 
not
surprising. At any distance into
the cube, the rays have a distribution of values for $p_{\rm K}$ for which 
typical values of
one standard deviation are indicated by the error brackets in Figure 1. Results 
for
computations in which the MHD fields (including variations in the density) are 
utilized
for the dense cube also are indicated in the panels for $N_{\rm corr}=12$ which
corresponds to the $N_{\rm corr}$ of the MHD simulations.

The variation of $p$ and $\sigma_\alpha$ in the dense cube is less than might be
expected for a simple model of the medium consisting of a number ($N_{\rm 
corr}$)
of independent cells along the path of the ray. This is partly because light 
that enters
the dense cube is already polarized. The contributions to the integral in
equation (1) from the diffuse and dense cubes will have opposite signs for
approximately half of the rays when, for example, there is no average magnetic 
field.
Such cancellations will tend to reduce the average increase of $p$. In
addition, for turbulent magnetic fields, the integrand in equation (1)
involves squares of three fields. It is quite irregular and is not well 
represented
by $N_{\rm corr}$ identical cells.

In Figure 2, we relate our calculations to the issue of the minimum $A_V$ at 
which
$(p/A_V)$ begins to be reduced in ISM clouds. The reduction seems to begin at
$A_V\approx1$ and the total extinction in these observations is
$A_V\approx4$ to 5. Hence, we adopt $A_V=1$ and 3 for the
diffuse and dense cubes, respectively, in the computations for Figure 1. Since 
the
observations are centered around 7500\AA\ (close to $V$), we adopt
$(p_V/A_V)_{\rm max} = 4$~\% mag$^{-1}$
(Mathis 1986) in equation (2). Results obtained when the MHD fields and density 
variations
are substituted in the dense cube are again shown in the Figure. Clearly, the 
calculations
are compatible with the observed variation of $\bar{p}_V$  with $A_V$ and
with the requirement that $\sigma_\alpha$ ($\approx20^\circ$)
change by no more than a few degrees for the starlight from the segment of the 
Taurus
cloud studied by Arce et al. (1998). As in Figure 1, at least several 
correlation lengths
across the dense cube and a $B_{\rm rms}$ that is somewhat greater than $B_{\rm 
avg}$
are required.

\section{Discussion}

At least within the idealized treatment here, the conclusion that irregularities 
in the
magnetic fields can allow the polarization characteristics $\bar{p}$ and 
$\sigma_\alpha$
to be relatively unaltered when starlight passes through the thick portion of 
ISM clouds
depends upon only two parameters in the clouds---$N_{\rm corr}$ and
$B_{\rm rms}/B_{\rm avg}$. Although we are not aware of
good information on $N_{\rm corr}$ in ISM clouds, $N_{\rm corr}\ga10$ is a 
reasonable
expectation (e.g., Myers \& Goodman 1991). From the Figures, $B_{\rm rms}/B_{\rm 
avg}$ must
then be greater than the ratio (1.5) for our weak field curve. In the MHD 
simulations with this ratio that we are utilizing, the corresponding rms 
turbulent velocity is about five
times the Alfv\'en velocity. In MHD
simulations more generally, the energy of the turbulent component of the 
magnetic field
is about 0.3 to 0.6 of the turbulent kinetic energy (V\'azquez-Semadeni et al. 
2000).
Super-Alfv\'enic turbulent velocities potentially have important implications 
for the
MHD structure of interstellar clouds as emphasized by Padoan \& Nordlund (1999) 
and later
by MacLow (2001).

For some years, the observation of polarized {\em emission} in the far-IR and
submillimeter has indicated
that dust grains must commonly be aligned in dense, optically thick ISM clouds
(e.g., Hildebrand 1996). It has, however,  seemed that selection effects might
reconcile the apparent conflict between these observations and the idea that the
reduced effectiveness of grains to create polarization in {\em extinction} is a 
result of the
failure of the alignment processes at higher optical depths. Observations of 
emission
have tended to select warm grains which occur
preferentially in environments with energy sources and are far from equilibrium, 
often in
the vicinity of young massive stars. In these environments, the alignment 
mechanisms
can be more effective (Lazarian, Goodman, \& Myers 1997). However, polarized
emission has recently been observed from dense, optically thick regions that 
also are
quiescent (Clemens, Kraemer, \& Ciardi 1999; Ward-Thompson et al. 2000; 
Vall\'ee,
Bastien, \& Greaves 2000). It seems difficult to understand that the processes 
which align
grains relative to the magnetic field should be more effective in these
relatively quiescent regions than in the clouds where the reduction in the
polarization caused by extinction is observed.

\acknowledgments

We are grateful to J. M. Stone for kindly making available results of the MHD
simulations, and to R. M. Crutcher for a number of valuable discussions. This 
research
has been supported by NSF Grants AST98-20641 and AST99-88104.

\clearpage

\begin{figure}
\plotone{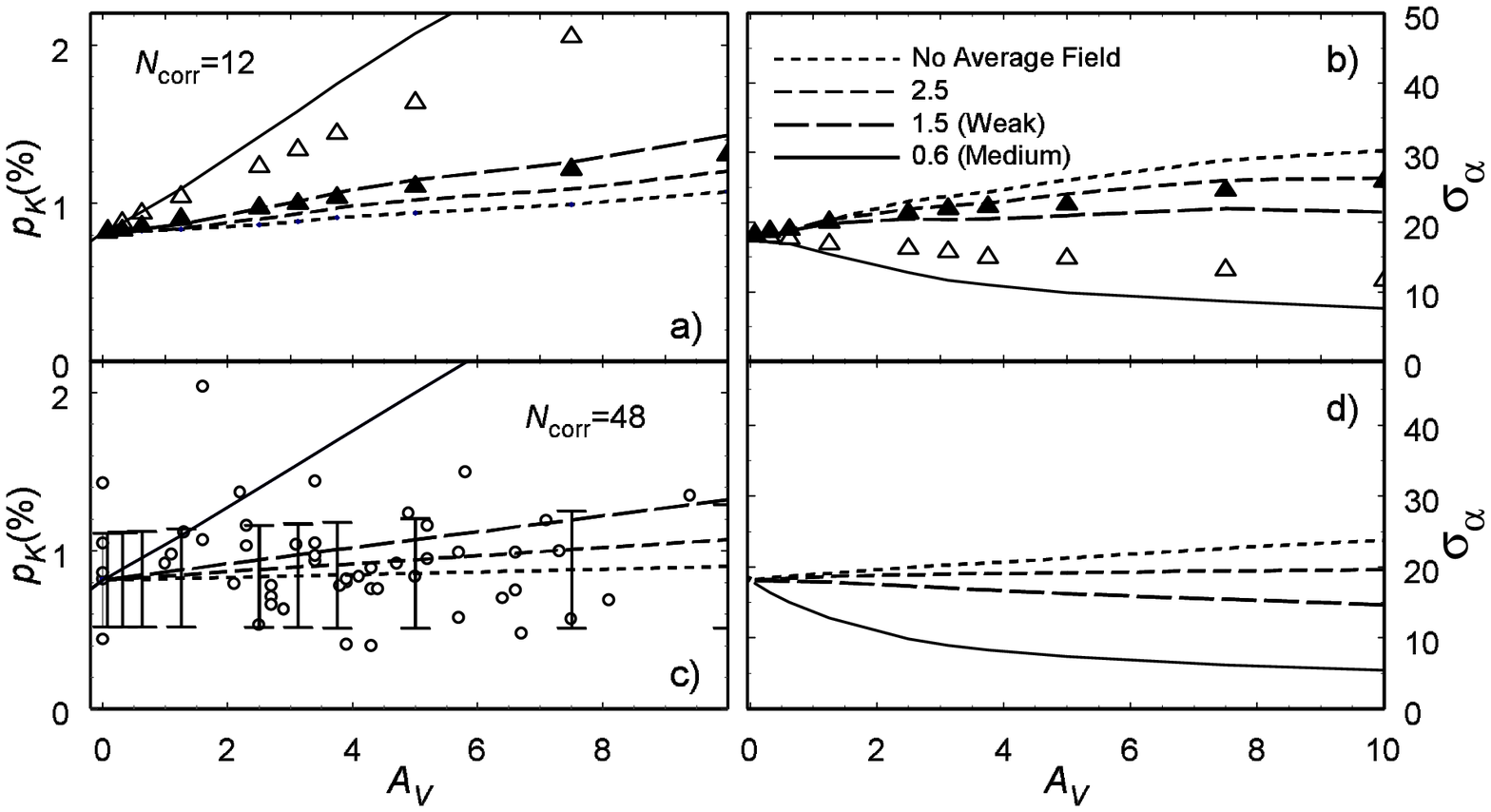}
\caption{%
(panels a and c) The fractional polarization in the IR (K band) averaged over 
all
rays $\bar{p}_{\rm K}$ versus optical extinction $A_V$ for starlight as it
propagates through a medium that represents the thicker part of an ISM cloud.
(panels b and d) The rms angle $\sigma_\alpha$ for the
dispersions in the position angles of this linear polarization. Calculations are 
performed
for several ratios $B_{\rm rms}/B_{\rm avg}$ of the irregular and average 
magnetic fields
as indicated by the various line types. The panels differ according to the 
number
of correlation lengths $N_{\rm corr}$ in the computational cube; $N_{\rm corr}= 
12$ for
panels a and b, and  $N_{\rm corr}= 48$ for panels c
and d.  Calculations also are performed with magnetic fields and variations in 
the matter
density obtained from the MHD simulations designated as ``medium'' and ``weak'' 
for
which  $B_{\rm rms}/B_{\rm avg}= 0.6$ and 1.5, respectively. These results are
indicated by open and filled triangles at several $A_V$
in panels a and b where $N_{\rm corr}= 12$ is similar to that of the MHD 
simulations.
The error brackets shown in panel c are representative for all curves in panels
a and c, and indicate one standard deviation.  Representative observational data
from L1755 (Goodman et al. 1995) are designated by open circles in panel c.
}
\end{figure}

\begin{figure}
\plotone{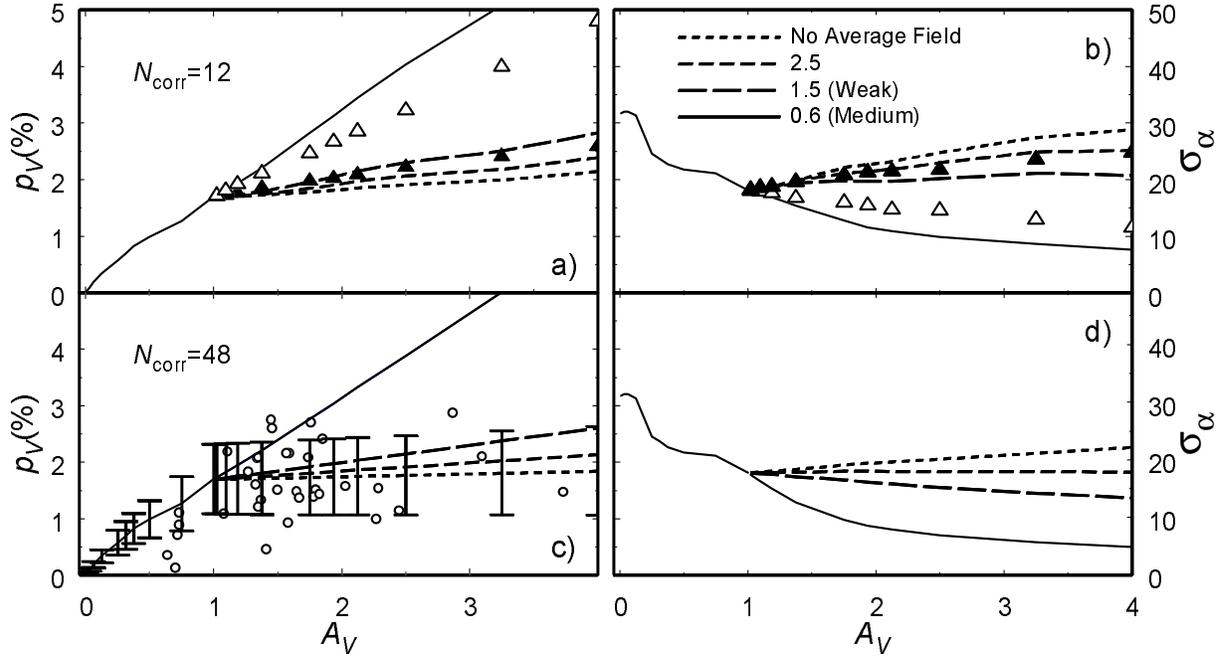}
\caption{%
Similar meaning to Figure 1, except that here $\bar{p}_V$ and $\sigma_\alpha$
are shown as the rays
of starlight first traverse a computational cubic volume which is intended to 
represent the
diffuse periphery of the cloud before traversing the second computational cube 
which
represents the thick part of the cloud. Representative observational data for
$\bar{p}_V$ from a region in Taurus (Arce et al. 1998) are indicated by open 
circles in 
panel c. Error brackets indicating one standard deviation, and which are 
representative for all
curves in panels a and c, are shown in panel c.
The results of calculations utilizing the MHD fields and variations in the 
matter density
are again designated by open and filled triangles in panels a and b.
}
\end{figure}

\end{document}